\documentclass[12pt]{iopart}
\usepackage{graphicx}  
\usepackage{psfig}  
\def\/{\over}
\def\<{\langle}
\def\>{\rangle}
\def\l<{\left\langle}
\def\r>{\right\rangle}
\def\({\left(}
\def\){\right)}
\def\[{\left[}
\def\]{\right]}

\def\mod{\,\mbox{mod}\,}

\def\d{{\rm d}}

\def\max{_{\rm max}}

\newcommand{\no}[1]{}

\def\sr{R}
\def\rc{r}
\def\yc{y_{\rm c}}
\def\reg{_{\rm reg}}
\def\ch{_{\rm ch}}
\def\uc{_{\rm uc}}
\def\ve{v_{0}}
\def\delx{\Delta x}
\def\dell{\Delta l}
\def\fig{.}
\begin{document}
\title[Directed chaos in magnetic billiards]
{Directed chaos in a billiard chain with transversal magnetic field}
\author{Holger Schanz and Manamohan Prusty}
\address{Max-Planck\--Institut f{\"u}r Dynamik und Selbstorganisation
  und\\ Institut f{\"u}r Nichtlineare Dynamik der Universit{\"a}t
G{\"o}ttingen\\
Bunsenstr. 10, D-37073 G{\"o}ttingen, Germany}
\ead{holger@chaos.gwdg.de}
\begin{abstract}
  In generic Hamiltonian systems with a mixed phase space chaotic transport
  may be directed and ballistic rather than diffusive. We investigate one
  particular model showing this behaviour, namely a spatially periodic
  billiard chain in which electrons move under the influence of a
  perpendicular magnetic field.  We analyze the phase-space structure and
  derive an explicit expression for the chaotic transport velocity. Unlike
  previous studies of directed chaos our model has a parameter regime in which
  the dispersion of an ensemble of chaotic trajectories around its moving
  center of mass is essentially diffusive. We explain how in this limit the
  deterministic chaos reduces to a biased random walk in a billiard with a
  rough surface. The diffusion constant for this simplified model is
  calculated analytically.
\end{abstract}
\section{Introduction}

With the term {\em directed chaos} we refer to extended chaotic systems in
which the time-averaged velocity of almost all chaotic trajectories approaches
for long time a non-zero constant $v\ch\ne 0$. In contrast to the well-known
deterministic chaotic diffusion, the chaotic transport is ballistic and
directed in such systems, $\<x\>=v\ch t$.  For 1D driven Hamiltonian systems
it has been shown that directed chaos may exist if the periodic driving is
such that (i) all generalized time-reversal symmetries are broken \cite{FYZ00}
and (ii) the system has a mixed phase space in which regular and chaotic
dynamics coexist \cite{S+01}. Examples of this type have been investigated in
a number of recent publications \cite{FYZ00,S+01,D+02,SDK05}, and closely
related phenomena were realized experimentally with cold atoms in pulsed
optical potentials \cite{M+02,J+05}.

In the present paper we address directed chaos in a different situation which
might be relevant in the context of solid-state physics. We study quasi
one-dimensional billiard chains with (i) a transversal magnetic field breaking
time-reversal symmetry and (ii) an asymmetric configuration of scatterers
which leads to a mixed phase space.  We will show that in these systems the
nature of transport is entirely different for the forward and the backward
direction. In one of the two directions transport is due to regular orbits
skipping along a boundary of the billiard chain (waveguide) without
back scattering. In the other direction there is strong back scattering and the
dynamics is chaotic. Consequently all dynamical properties such as the average
transport velocity or the superimposed spreading of a distribution of
particles are different for the two transport directions. It is this special
property which makes the systems we consider interesting and potentially
useful for controlling nanoscale electronic transport.

We can considerably extend a preceeding study of directed chaos in magnetic
billiards \cite{AD03} because the specific geometry which we propose leads to
a particularly simple phase-space structure. This allows for detailed
analytical calculations and provides sufficient insight in order to control
the transport properties of our model with a few geometric parameters. For
example, we obtain an explicit expression for the chaotic transport velocity.
We can also understand in some detail the velocity dispersion of the chaotic
trajectories. In known examples for directed chaos
\cite{FYZ00,S+01,D+02,SDK05} this dispersion was found to be non-Gaussian,
with clear signatures of L\'evy walks. In contrast, we identify a parameter
that can be tuned such that the dispersion becomes essentially diffusive while
$v\ch\ne 0$ remains constant. In other words, there really is a clear
separation of scales between directed transport, $\<x\>\sim t$, and undirected
broadening of an ensemble, $\<\Delta x\>\sim t^{1/2}$.  This might be a
desirable feature in applications.  In this regime of biased diffusion the
chaotic dynamics can be approximated by a non-deterministic random walk in a
billiard with a rough boundary, although our original model is deterministic
and has no disorder. Rough billiards are nowadays standard models for electron
dynamics on mesoscopic scales \cite{FS97b,LFL98}.  However, the system which
we study here seems to be the first such model with a {\em mixed
  regular-random} phase space and is thus interesting also in its own right.

In the following two main sections we introduce and analyze our billiard
models with directed chaos.  Sec.~\ref{sec:sinai} is devoted to a periodic
billiard chain with transversal magnetic field in which the dynamics is
deterministic, while in Sec.~\ref{sec:rough} we study a billiard with a rough
surface and non-deterministic dynamics. The connection between these two
models is explained in Sec.~\ref{sec:disperse}. Sec.~\ref{sec:sum} closes our
paper with a summary and some concluding remarks.

\section{Periodic chain of magnetic Sinai billiards}\label{sec:sinai}

\begin{figure}[htb]
 \centerline{\psfig{figure=\fig/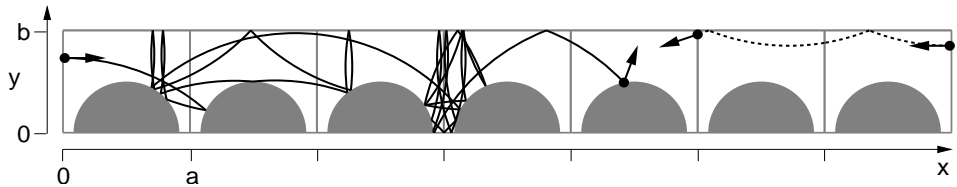,width=13cm}}
 \caption{\label{billiard} 
   In a waveguide with obstacles placed only along one of the walls a
   transversal magnetic field leads to the coexistence of regular and chaotic
   trajectories. Two corresponding examples are shown with a dashed and a full
   line, respectively. Note that for the same magnetic field and particle
   charge the trajectories can be traversed only with the indicated
   orientation. The relevant parameters of the model are $a=1.2b$ (spatial
   period), $\sr=0.5b$ (radius of obstacles) and $\rc=2b$ (cyclotron radius).}
\end{figure}

In Fig.~\ref{billiard} we show an example for an asymmetric magnetic billiard
which illustrates qualitatively the effect we investigate. The system consists
of a waveguide in which one wall is perfectly straight, $y_{1}(x)\equiv b$.
The opposite wall is distorted periodically, $y=y_{2}(x)\ge 0$,
with
\begin{equation}
y\max={\rm max}_{x}\,y_{2}(x)<b\,.
\end{equation}
Specifically in our case this is achieved by a
chain of semi-circular obstacles with radius $\sr=y\max$, but other geometries
may lead to similar results\footnote{For example, we found similar behavior when
  rectangular unit cells are extended by attaching on one side a
  semicircle instead of cutting it out \cite{Pru06}.}.

Energy is conserved, and we will consider the dynamics on the energy shell
corresponding to the velocity $\ve$. For numerical calculations we use
dimensionless units in which $\ve=b=1$, i.e., we measure velocity in units of
$\ve$, lengths in units of $b$ and time in units of $b/\ve$. As a consequence,
the geometric length of a trajectory is equal to the elapsed time. Due to a
perpendicular magnetic field the trajectories of electrons in the waveguide
consist of circular arcs with cyclotron radius $\rc=m\ve/eB$. In the following
we will use the cyclotron radius to parameterize the magnetic field.

The magnetic field leads to a special set of trajectories skipping along the
clean wall of the channel (dashed line in Fig.~\ref{billiard}). It is a matter
of simple geometry to describe the transport due to these regular trajectories
which by our convention of the magnetic field is directed to the left
(negative transport velocity).  In contrast, trajectories colliding with the
distorted wall of the waveguide can be chaotic such as the example shown with
a full line in Fig.~\ref{billiard}. The figure suggests that such trajectories
are transporting to the right, that is opposite to the skipping ones. We will
see below that this is indeed the case.  This system shows directed chaos, and
the transport due to the chaotic trajectories compensates for the regular
skipping orbits such that no net transport results if one averages over all
possible initial conditions.

\subsection{Phase space structure}

A phase-space point on the energy shell is completely described by the
position $(x,y)$ in the waveguide and the angle $\varphi$
between the velocity vector and the $x$-axis.  In these coordinates, the invariant measure is
\begin{equation}\label{mu}
\d\mu=\d x\,\d y\,\d\varphi\,.
\end{equation}
\begin{figure}[htb]
 \centerline{\psfig{figure=\fig/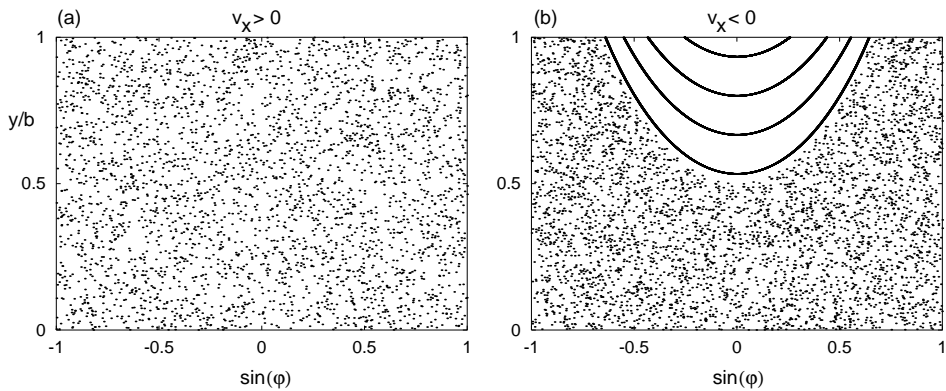,width=13cm}}
 \caption{\label{psec} Phase space portrait of the system of 
   Fig.~\protect\ref{billiard}. $y$ and $\varphi$ are shown for $x=0\mod a$
   and (a) $\cos\varphi>0$, (b) $\cos\varphi<0$. The data represent one
   chaotic and four skipping trajectories.}
\end{figure}
In order to obtain a general impression of the dynamical scenarios supported
by our model we show a Poincar\'e section (Fig.~\ref{psec}) obtained by
marking $y$ and $\varphi$ whenever a trajectory crosses the boundary of a unit
cell, $x=0\mod a$, i.e., whenever $x$ is half way between two obstacles. The
two panels correspond to crossings with positive and negative transport
velocity
\begin{equation}
v_{x}=\ve\cos\varphi\,.
\end{equation}
Note that the invariant measure of the resulting Poincar\'e map is
\begin{equation}\label{muper}
\d\mu_{\perp}=\d y\,\d\varphi\,\cos\varphi\,.
\end{equation}
In the Poincar\'e section we observe a big chaotic component which contributes
to both directions. In fact the randomly scattered points in Fig.~\ref{psec}a
and b are the trace of a single chaotic trajectory, namely the one shown in
Fig.~\ref{billiard}.  In addition we see in Fig.~\ref{psec}b also the traces
of a number of skipping orbits.  These trajectories have a second constant of
the motion besides the energy, namely the $y$-component of the current arc's
center
\begin{equation}
\yc(y,\varphi)=y-\rc\cos\varphi\,.
\end{equation}
Therefore the skipping orbits constitute a regular (integrable) component of
phase space which comprises all phase-space points satisfying
\begin{equation}\label{skipcond}
\cos\varphi\le {y-\sr\over \rc}-1\,,
\end{equation}
that is all arcs whose lowest point is still above the top of the obstacles,
$\yc(y,\varphi)-\rc\ge y\max$.  It is clear that trajectories with such
initial conditions can never be scattered. One may argue that there can exist
also skipping orbits violating the condition in Eq.~(\ref{skipcond}). Say the
lowest point of the current arc is below $y\max$ but falls into the gap
between two obstacles. However, if one continues the trajectory it will
finally hit one of the scatterers, at least if the longitudinal extension
$\Delta x$ of the arcs is not rationally related to the spatial period $a$.
Thus, exceptional skipping orbits within the chaotic component may exist but
must have measure zero in phase space.

Are all orbits either chaotic or skipping? There is no reason to expect this.
Typically, the boundary between a regular and a chaotic component in phase
space is fractal, with hierarchies of smaller and smaller stable islands
embedded into the chaotic sea. Also deep within the chaotic sea stable
periodic orbits with small islands surrounding them can exist. Such islands
are clearly visible for relatively small cyclotron radius, e.~g.\ $\rc=1$.
However, for the parameters used in Figs.~\ref{billiard} and \ref{psec} we
were unable to detect stable islands even after increasing the resolution of
the phase space portrait considerably. For growing cyclotron radius it must be
expected that stable islands are less and less important, as for zero magnetic
field ($\rc=\infty$) the system reduces to the Sinai billiard (Lorentz gas)
which is known to be fully hyperbolic \cite{Sin70}. Hence, we will neglect the
influence of additional stable islands and assume that for sufficiently large
$\rc$ the phase space of our model has the simple structure which is visible
in Fig.~\ref{psec}. All analytical calculations below will be based on this
assumption.

\subsection{Average chaotic transport velocity}

\begin{figure}[htb]
 \centerline{\psfig{figure=\fig/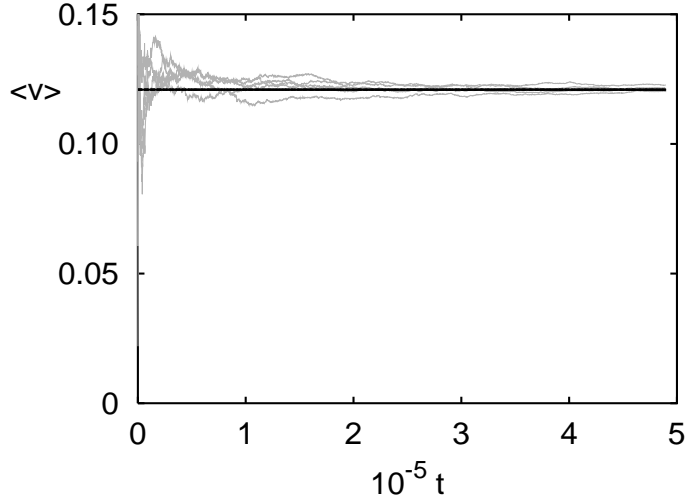,width=10cm}}
 \caption{\label{vmean} For the parameters of Fig.~\ref{billiard}
   the time-averaged velocity of 5 different chaotic trajectories is shown as
   a function of time. For all trajectories the average velocity approaches
   the value predicted by Eq.~(\ref{vch}) which is indicated by the horizontal
   line.}
\end{figure}

We will now calculate the asymptotic chaotic transport velocity. For this
purpose it is useful to restrict the dynamics to a unit cell by taking $x$
modulo $a$. After this transformation the phase-space volume of the system is
finite but the transport velocity $v_{x}(t)=\ve\cos\varphi(t)$ is unchanged.
Assuming ergodicity within the chaotic component of the unit cell, the
long-time average over the transport velocity converges for almost all chaotic
trajectories to the corresponding phase-space average,
\begin{eqnarray}\label{ergo}
v\ch&=&\lim_{t\to\infty}t^{-1}\int_{0}^{t}\d t'\,v_{x}(t')
\nonumber\\&=&
\ve\,\<\cos\varphi\>\ch\,.
\end{eqnarray}
This is illustrated in Fig.~\ref{vmean}. As common in
systems with a mixed phase space, the assumption of ergodicity within the 
chaotic component cannot be justified analytically. Numerical evidence in
favour of it has been gathered for our model by checking that individual
chaotic trajectories cover the available phase-space region uniformly.

In general there is no reason to expect $v\ch=0$ in Eq.~(\ref{ergo}). However,
an average over $\cos\varphi$ is obviously zero, if for any point
$(x,y,\varphi)$ in the domain of integration also the point $(x,y,\varphi+\pi)$
with opposite velocity direction contributes. For example this is the case 
if the integration extends over the entire phase space of the unit cell,
\begin{equation}\label{unbiased}
\<\cos\varphi\>\uc=0\,.
\end{equation}
Hence, in a completely chaotic billiard we would have $v\ch=0$, i.e., there is
no directed chaos. Similarly, in any system with a single chaotic component
and time-reversal symmetry there can be no directed chaos even if stable
islands are present.  On the other hand directed chaos can exist in fully
chaotic systems with time-reversal symmetry if the phase space contains two
invariant chaotic components, see \cite{HP04} for an example.

In our case we can decompose the full phase space into the chaotic component and the
skipping orbits. Thus we have
\begin{equation}
\Omega\uc\<\cos\varphi\>\uc=\Omega\reg\<\cos\varphi\>\reg+\Omega\ch\<\cos\varphi\>\ch
\end{equation}
and  
\begin{equation}
\Omega\uc=\Omega\reg+\Omega\ch\,,
\end{equation}
where $\Omega_{i}=\int_{i}\d\mu$ denotes the phase-space volume of the set $i$. Substitution into
Eq.~(\ref{ergo}) yields
\begin{equation}\label{vchom}
v\ch=-\ve{\Omega\reg\,\<\cos\varphi\>\reg\over \Omega\uc-\Omega\reg}
\end{equation}
which is the analogue to the sum rule derived in \cite{S+01} for the
chaotic transport velocity in driven 1D systems. However, in the present case
we are able to calculate $v\ch$ explicitly. First we note that the phase-space
volume of the unit cell is 
\begin{eqnarray}
\Omega\uc&=&2\pi\(ab-{\pi\over 2}\sr^2\)\,.
\end{eqnarray}
Next we introduce the characteristic function of the regular component which,
according to Eq.~(\ref{skipcond}), is given by
\begin{equation}
\chi\reg(x,y,\varphi)=\Theta\({y-\sr-\rc\over \rc}-\cos\varphi\)\,.
\end{equation}
The volume of the regular component is then
\begin{eqnarray}
\Omega\reg&=&\int\d\mu\,\chi\reg(x,y,\varphi)
\nonumber\\&=&
a\int_{0}^{2\pi}\d\varphi\,
\int_{0}^{b}\d y\;
\Theta\!\({y-\sr-\rc\over \rc}-\cos\varphi\)
\nonumber\\&=&
a\int_{\pi-\theta}^{\pi+\theta}\d\varphi\,
(b-[\sr+\rc+\rc\cos\varphi])
\nonumber\\&=&2a\rc(\sin\theta-\theta\,\cos\theta)
\end{eqnarray}
where 
\begin{equation}\label{theta}
\theta=\arccos\(1-{b-\sr\over \rc}\)
\end{equation}
denotes the maximum deviation of the limiting skipping orbit from the horizontal. Similarly we have
\begin{eqnarray}
\Omega\reg\<\cos\varphi\>\reg&=&\int\d\mu\,\chi\reg(x,y,\varphi)\,\cos\varphi
\nonumber\\&=&a\int_{\pi-\theta}^{\pi+\theta}\d\varphi\,
(b-[\sr+\rc+\rc\cos\varphi])\,\cos\varphi
\nonumber\\&=&-a\rc\(\theta-\sin\theta\cos\theta\)\,.
\end{eqnarray}
Substitution of these expressions into Eq.~(\ref{vchom}) finally yields
\begin{equation}\label{vch}
v\ch={\ve\/2}{\theta-\sin\theta\cos\theta\/(\pi/a\rc)\(ab-{\pi\over 2}\sr^2\)-
(\sin\theta-\theta\,\cos\theta)}
\end{equation}
for the chaotic transport velocity. This result is confirmed numerically in
Figs.~\ref{vmean} and~\ref{diffuse}.

For small magnetic field, when our assumption of a two-component phase space
is justified best, we can approximate Eq.~(\ref{vch}) by the leading order in the
cyclotron radius. We find 
\begin{equation}
\theta\approx\sqrt{2(b-\sr)\/\rc}
\end{equation}
and
\begin{equation}
v\ch\approx\ve{4a\/3\Omega\uc}\sqrt{2(b-\sr)^{3}\/\rc}\,.
\end{equation}
This shows explicitly that the chaotic transport vanishes in the absence of a
magnetic field ($\rc\to\infty$). 

\subsection{Dispersion of chaotic trajectories}\label{sec:disperse}

With the above results we can precisely predict the average transport velocity
of many chaotic trajectories at a given moment in time, or the long-time
average over a typical trajectory. However, we have no precise information
about the dispersion of an ensemble, or the rate of convergence of the
transport velocity to its asymptotic value. In systems with sufficiently fast
decay of correlations, the dispersion of chaotic trajectories is diffusive,
\begin{equation}
\<[x(t)-v\ch t]^2\>=Dt\,.
\end{equation}
In this case the distribution of time-averaged velocities
\begin{equation}
\overline v_{t}={x(t)-x(0)\/t}
\end{equation}
is for long times a Gaussian with variance $D/t$,
\begin{equation}
P(\overline v_{t})=\sqrt{t\/2\pi D}\exp\(-[t/2D]\,[\overline v_{t}-v\ch]^2\)\,.
\end{equation}
\begin{figure}[t]
 \centerline{\footnotesize\sf
\begin{tabular}{ll}(a)&(b)\\[-8mm]
  \psfig{figure=\fig/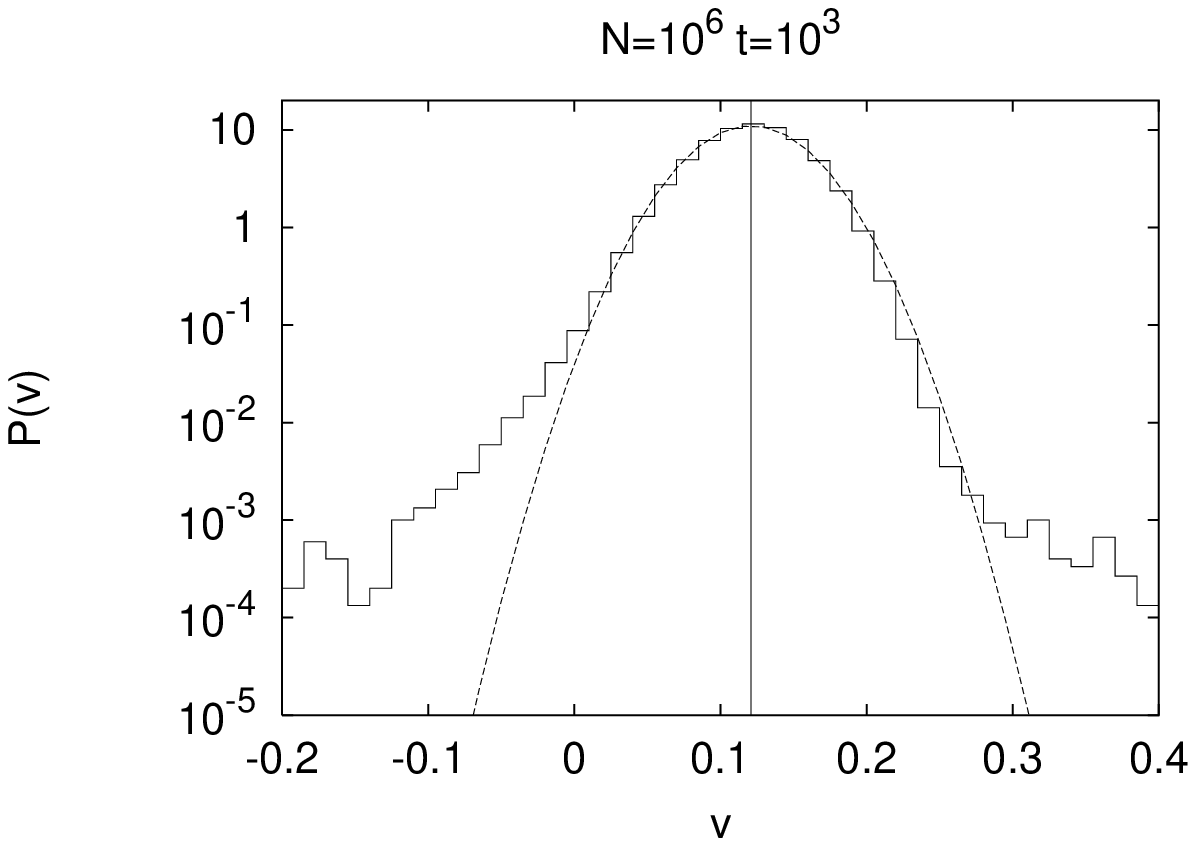,width=6.5cm}&
  \psfig{figure=\fig/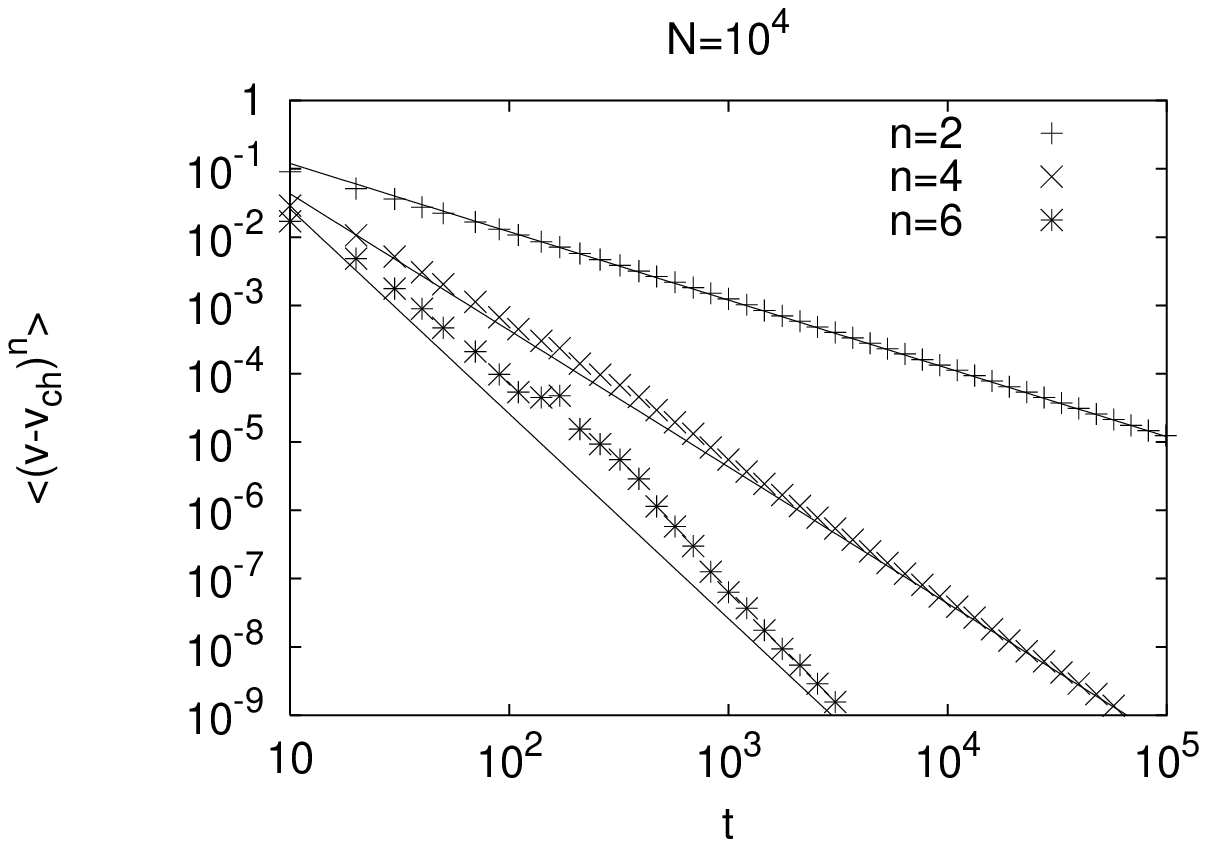,width=6.5cm}
\end{tabular}
}
 \caption{\label{diffuse} For the model of Fig.~\ref{billiard} the dispersion
   of chaotic transport velocities is analyzed. (a) $N=10^6$ trajectories have
   been iterated up to $t=10^3$. The resulting distribution of average
   velocities is displayed as a histogram and compared to a Gaussian with the
   same mean and variance (dotted line). The prediction of Eq.~(\ref{vch}) for the
   average velocity is indicated by a vertical line. The numerical mean value
   of the distribution differs from that by $1.5\times10^{-5}$. (b) shows some
   higher moments $\<(\overline v-v\ch)^n\>$ of the distribution as a function
   of time. The straight lines indicate the corresponding moments of a
   diffusively spreading Gaussian with $D=1.2$.}
\end{figure}
\begin{figure}[t]
 \centerline{\footnotesize\sf
\begin{tabular}{ll}(a)&(b)\\[-8mm]
  \psfig{figure=\fig/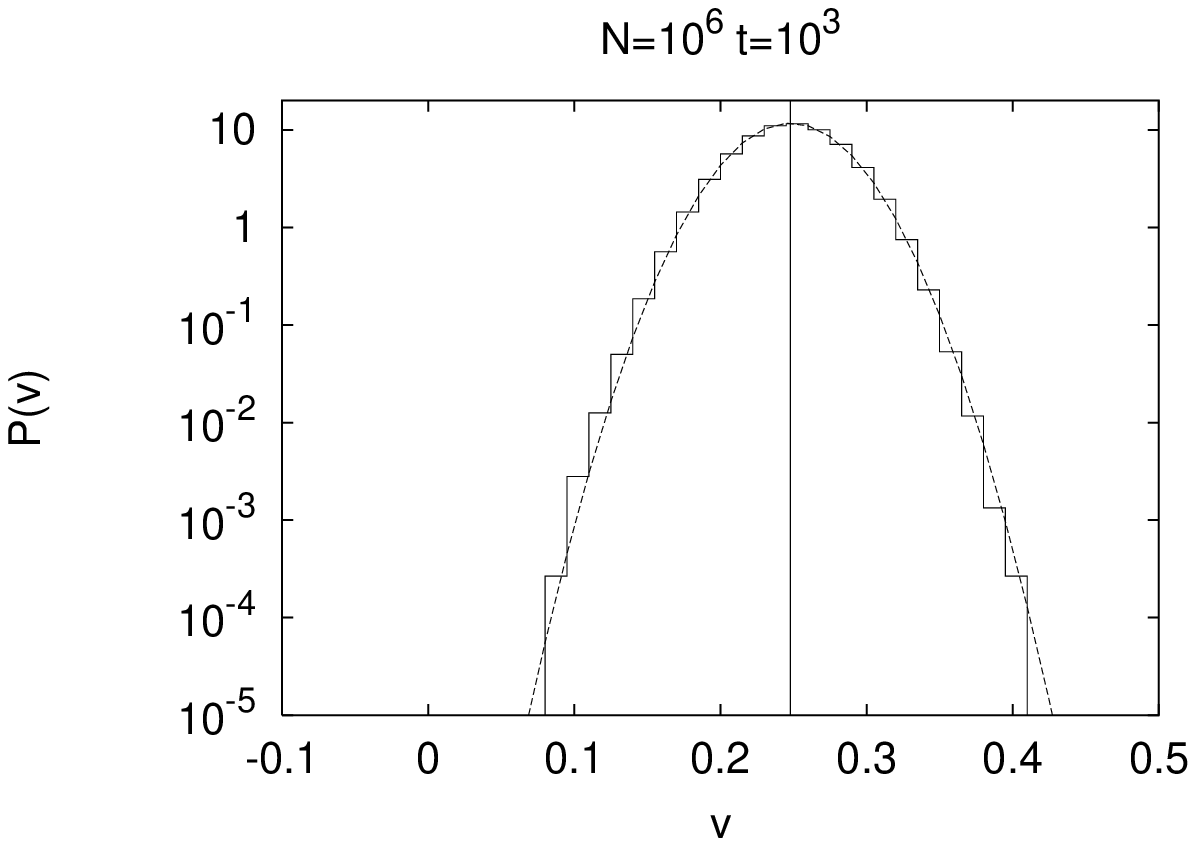,width=6.5cm}&
  \psfig{figure=\fig/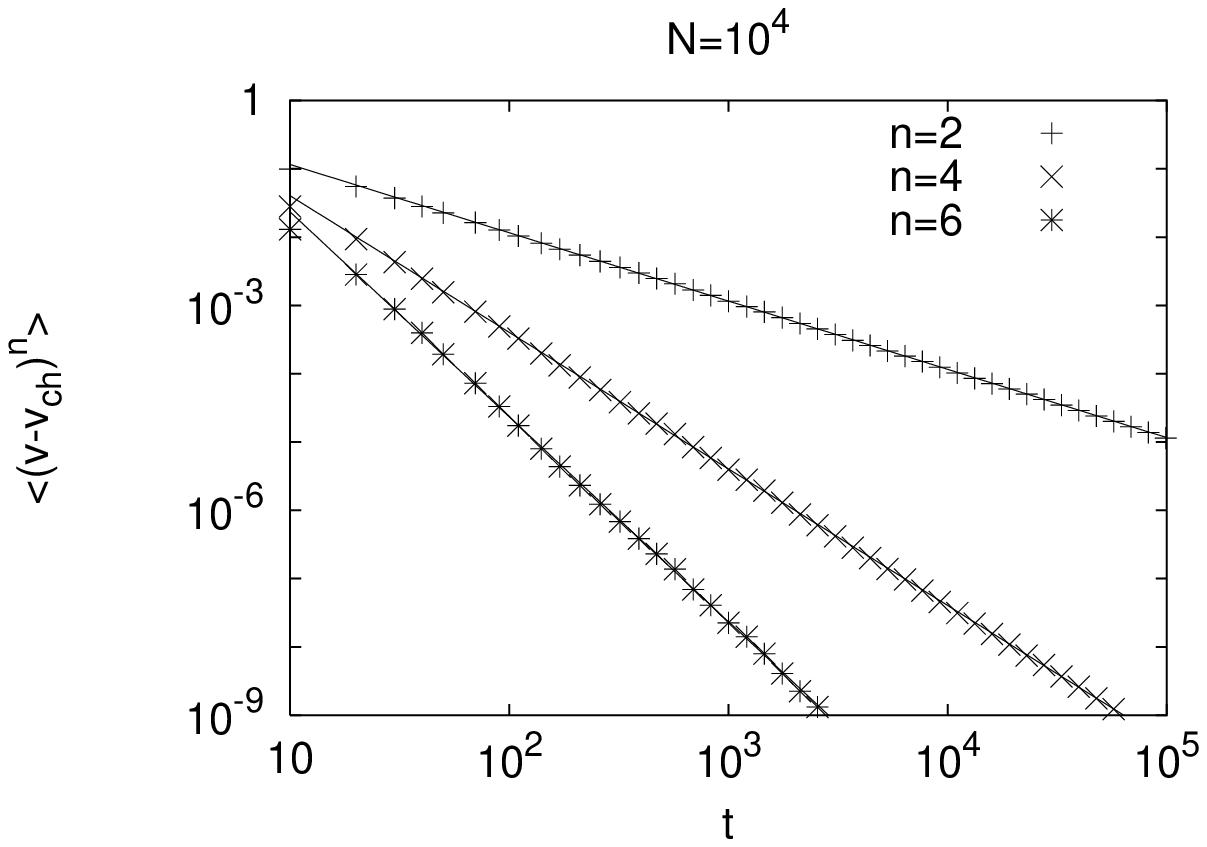,width=6.5cm}
\end{tabular}
}
 \caption{\label{diffuse2} Same as Fig.~\ref{diffuse} for $\sr=0.01b$ and
  $a=2\sr$ (neighbouring semicircles are touching). The fitted diffusion
  constant is $D=1.2$ }
\end{figure}
In Fig.~\ref{diffuse} we compare this hypothesis to the chaotic dispersion
for the parameters of Fig.~\ref{billiard}. Clear deviations from diffusive
behavior are observed in the tails of the distribution and for its higher
moments. After iterating $N=10^6$ trajectories to $t=10^3$ the average chaotic
velocity agrees very well with the value predicted by Eq.~(\ref{vch}).
However, the distribution of time-averaged velocities is asymmetric around
$v\ch$  and, compared to a Gaussian, large deviations from the mean value
are enhanced. For the considered time interval the variance of the
distribution (+) is fitted well by the diffusive prediction $D/t$. However,
with increasing $n$ the higher moments $\<(v-v\ch)^n\>$ of the velocity
distribution deviate more and more from the corresponding time-dependent
Gaussian ($\times$, $*$ vs full lines).  This is no surprise as in systems with a mixed phase
space anomalous diffusion is typical.  Indeed, in previous studies of directed
chaos in such systems the observed velocity distributions were always
non-Gaussian \cite{FYZ00,S+01,D+02,SDK05}. They displayed traces of L\'evy walks and
anomalous diffusion which were even stronger than those in Fig.~\ref{diffuse},
and it was not possible to adjust the model parameters such that directed
transport coexists with normal diffusion.

\begin{figure}[tb]
 \centerline{
  \psfig{figure=\fig/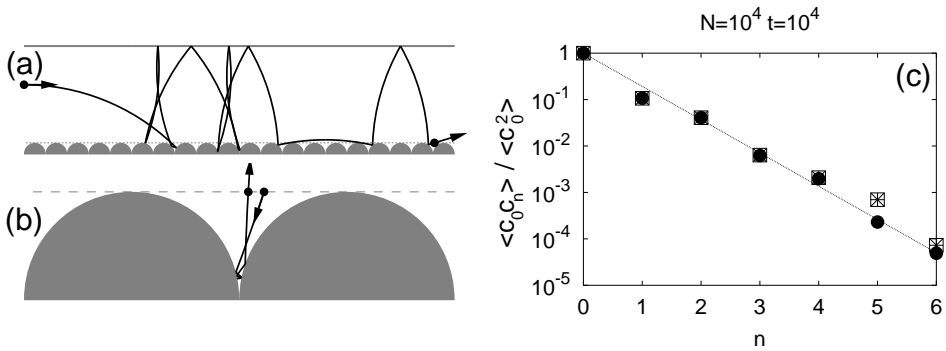,width=13cm}
 }
 \caption{\label{smallball} (a) A piece of a chaotic trajectory is shown for $\sr=0.1b$ 
   and $a=2\sr$. (b) One particular scattering of this trajectory from the
   lower billiard boundary is magnified. Note that on this scale the curvature
   of the trajectory due to the magnetic field is hardly visible. (c) The
   normalized correlation between consecutive intersection angles of the segments
   with the line $y=\sr$ is shown with stars for the model of (a), with empty
   squares for $\sr=0.01b$ and and with black dots for the same $\sr$ but
   randomized $\xi_n$ (see text). The straight line shows the fitted
   exponential $\exp(-1.65 n)$.}
\end{figure}

Therefore it is quite remarkable that in our model a slightly modified
geometry leads to a velocity distribution which numerically could not be
distinguished from a Gaussian. In Fig.~\ref{diffuse2} the analysis of
Fig.~\ref{diffuse} is repeated for a system in which (i) the semicircular
obstacles are much smaller than in Fig.~\ref{billiard} and (ii) neighbouring
scatterers are tangent rather than separated by a gap (see
Fig.~\ref{smallball}a). 

\begin{figure}[t]
 \centerline{
  \psfig{figure=\fig/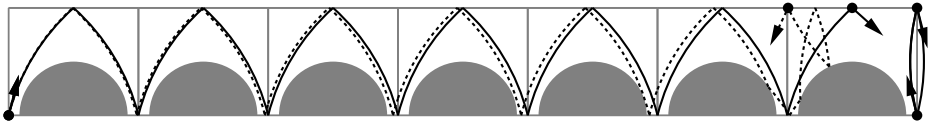,width=13cm}
 }
 \caption{\label{margstab} The full lines show two marginally stable periodic
   orbits (one of them is periodic up to translation by one unit-cell length).
   The dotted line is a non-periodic chaotic trajectory which remains in the
   vicinity of a marginally stable orbit for a very long period of time.}
\end{figure}

This latter modification removes marginally stable periodic orbits which hit
the lower wall within the straight sections of the boundary only
(Fig.~\ref{margstab}). Up to a certain limit it is possible to translate such
orbits along the channel without changing their shape.  Nevertheless they form
a set of measure zero in phase space since the angle of the trajectory cannot
be varied without destroying the orbit (dotted line in Fig.~\ref{margstab}).
It is known that marginally stable trajectories are a source of
non-exponential decay of correlations. They lead to anomalous diffusion even
in completely chaotic systems like the Sinai or the stadium billiard. In our
system generic chaotic trajectories can remain in their vicinity for a long
time and during these episodes keep the transport velocity substantially above
or below the average (e.~g.\ $v\approx 0.5$ for the dotted line in
Fig.~\ref{margstab} while $v\ch=0.12$). Therefore it is quite natural that the
chaotic dispersion shows much less anomalous diffusion when the specific
geometry prevents the existence of marginally stable trajectories.

In contrast, it is not immediately clear why reducing the radius of the
scatterers should further suppress anomalous diffusion in our model. In order
to find an intuitive explanation for this numerical fact we consider now a
different Poincar\'e section defined by the intersections of a trajectory with the line
$y=\sr$.  Note that regular skipping orbits do not at all intersect this line.
In analogy to Eq.~(\ref{muper}) the invariant measure restricted to the
surface of section is now
\begin{equation}\label{mupar}
\d\mu_{\parallel}=\d x\,\d\varphi\,\sin\varphi=\d x\,\d c
\end{equation}
with $c=\cos\varphi$. Hence, a typical chaotic trajectory is represented in
this Poincar\'e section by points $(\xi_{n},c_{n})$ which uniformly cover the
area $[0,1]\times[-1,1]$.  $\xi_{n}=(x_{n}/a)\mod 1$ denotes here the position
of the intersection relative to the unit cell. Following a transition across
$y=\sr$ from below, $(\xi_{n-1},c_{n-1})$, there will always be an
intersection from above which is denoted by $(\xi_{n}',c_{n}')$. Geometry
requires $c_{n}'=c_{n-1}$.

For small obstacle radius transport is mainly due to the long segments of the
trajectory above $y=R$ (Fig.~\ref{smallball}a). Each segment is either a
single circular arc or a combination of two arcs which are identical up to a
reflection. Segment $n$ is entirely characterized by the value of $c_{n}$.
Therefore the statistics of the $c_{n}$ determine the transport properties. In
particular, the velocity dispersion will be diffusive if correlations between
consecutive angles decay sufficiently fast.  This argument will be made more
explicit in Sec.~\ref{sec:rough} below.  Fig.~\ref{smallball}c shows that for
sufficiently small obstacle radius $\<c_{0}c_{n}\>$ indeed decays very fast as
a function of $n$ (exponentially fast to a good approximation).  This
observation can be understood as follows: As $\sr\to 0$, a segment
$(\xi_{n-1},c_{n-1})\to(\xi_{n}',c_{n}')$ traverses more and more unit cells
between its end points. As a consequence, one expects an extremely rapid
variation of $\xi_{n}'$ as a function of $c_{n-1}$. This suggests to replace
$\xi_{n}'$ by a random variable.  Fig.~\ref{smallball}c confirms that this
approximation is very good for the parameters of Fig.~\ref{smallball}a or even
smaller obstacle radius $\sr$.  Note that the argument leading to this
randomization does not depend on the value of the cyclotron radius $\rc$. In
particular, it remains valid also in the absence of a magnetic field,
$\rc=\infty$.

Next we consider the dynamics in the vicinity of the lower boundary, i.e., for
$y<R$. For small obstacle radius it is practically identical to that for
$\rc=\infty$ since the ratio $2\sr/\rc$ between the size of the unit cell and
the (constant) radius of curvature of the trajectory vanishes as $\sr\to 0$
(Fig.~\ref{smallball}b). Hence the resulting mapping
$(\xi_{n},c_{n})\to(\xi_{n}',c_{n}')$ may be approximated by a Sinai billiard
without magnetic field. The latter is known to be hyperbolic \cite{Sin70} and
shows exponential mixing if no marginally stable orbits are present. As the
correlations should not be enhanced by the additional randomization of $\xi$
we finally understand why $\<c_{0}c_{n}\>$ decays exponentially for small
obstacle radius $\sr\ll\rc$, leading to a diffusive  velocity dispersion.

\section{A waveguide with one rough boundary}\label{sec:rough}

The results of the previous section suggest that for small obstacle radius
$\sr$ the detailed dynamics inside the lower boundary layer of the waveguide
is irrelevant. This holds at least as long as the dynamics therein is chaotic.
In this case the essential physical properties of our model are not affected
if we replace the boundary layer by an idealized rough surface and describe
the scattering at the lower wall probabilistically. In the present section we
follow this approach and obtain in this way a better analytical understanding.
For the sake of simplicity we will continue to refer to trajectories which hit
the lower wall as {\em chaotic} although the dynamics is not deterministic
anymore. This is justified since we have just explained how random scattering
may arise from a deterministic chaotic system in a certain limit. Nevertheless
there are also other mechanisms such as thermal or quantum
fluctuations which may lead to the same probabilistic model.

\begin{figure}[tb]
 \centerline{
  \psfig{figure=\fig/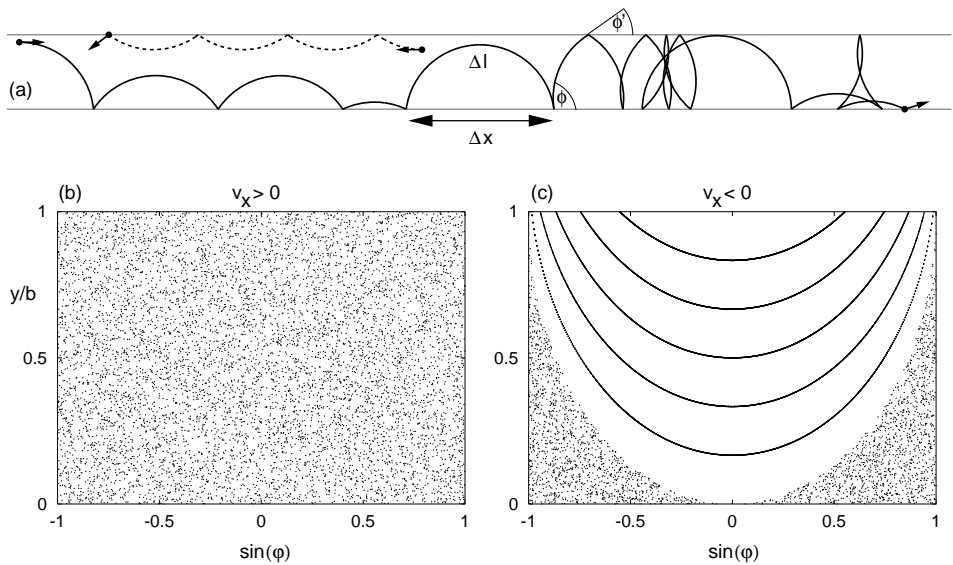,width=13cm}
 }
 \caption{\label{rough} (a) Typical electron trajectories (dashed line:
   regular, full line: chaotic) for a waveguide with a perpendicular magnetic
   field ($\rc=5b/3$). The lower wall is perfectly rough ($\alpha=0$) such
   that the angles before and after a reflection are completely uncorrelated.
   (b) and (c) show a phase space portrait of this system for $r=b$.}
\end{figure}

In order to define our simplified model we must prescribe the probability for
a trajectory to leave the lower boundary at an angle $0\le\phi\le\pi$.
Consistency with the invariant measure $d\mu_{\parallel}$, Eq.~(\ref{mupar}),
requires
\begin{equation}\label{pphi}
P(\phi)={\sin\phi\over 2}\qquad(0\le \phi\le\pi)\,.
\end{equation}
In terms of $c=\cos\phi$ this amounts to
\begin{equation}\label{pc}
P(c)={1\over 2}\qquad(-1\le c\le +1)
\end{equation}
for the probability density averaged over all chaotic trajectories. We account
for possible correlations between the segments of a trajectory by
introducing a probability $\alpha$ for a specular reflection at the lower
boundary. The probability density for $c_{n}$ is then
\begin{equation}\label{pcn}
P(c_{n})=\alpha\,\delta(c_{n}-c_{n-1})+{1-\alpha\/2}\,.
\end{equation}
While for $\alpha=0$ there are no correlations between consecutive values of
$c_{n}$
\begin{eqnarray}
\<c_{0}c_{n}\>&=&\delta_{n,0}\int_{-1}^{+1}\d c\,P(c)\,c^2
\nonumber\\&=&{\delta_{n,0}\/3}\qquad\qquad(\alpha=0)\,,
\end{eqnarray}
the possibility of specular reflections leads to exponentially decaying correlations 
\begin{equation}
\<c_{0}c_{n}\>={\alpha^{n}\/ 3}\qquad(\alpha\ne0)\,.
\end{equation}
For each segment of a chaotic trajectory we can express the horizontal distance
between its endpoints as
\begin{eqnarray}\label{delx}
\delx(\phi)&=&2\rc(\sin\phi-\sin\phi')\,.
\end{eqnarray}
On the other hand the total length of a segment is
\begin{eqnarray}\label{dell}
\dell(\phi)&=&2\rc\,(\phi-\phi')\,.
\end{eqnarray}
Here $\phi$ and $\phi'$ denote the angles of intersection with the lower and the
upper wall of the channel, respectively (see Fig.~\ref{rough}a). In terms of
the former, the latter is given by
\begin{eqnarray}
\phi'=\left\{\begin{array}{lr}
\arccos(\cos\phi+{b\/\rc})\hspace*{5mm}&(\cos\phi\le 1-{b\/\rc})
\\[2mm]
0&(\cos\phi> 1-{b\/\rc})
\end{array}\right.\,,
\end{eqnarray}
where the second line extends the definition of $\phi'$ to those segments of a
trajectory which do not reach the upper wall.

With the help of the angle $\phi$ we can give an alternative parameterization of
the chaotic component of phase space. Instead of $(x,y,\varphi)$ we can refer
to one of its points by the coordinates $(x,\phi,l)$. As mentioned above,
$\phi$ is the inclination angle of the segment at its initial point, and $0\le
l\le \Delta l(\phi)$ is the path length measured from there. Formally, the
transformation between $(\phi,l)$ and $(\varphi,y)$ is given by
\begin{eqnarray}\label{vt1}
\varphi(\phi,l)&=& 
\left\{\begin{array}{ll}
\phi-{l\/\rc} &l\le{\Delta l(\phi)\/2} \\[3mm]
-\varphi(\phi,\Delta l-l)\qquad & l>{\Delta l(\phi)\/2}
\end{array} \right. 
\\\label{vt2}
y(\phi,l)&=&\rc\cos\varphi(\phi,l)-\rc\cos\phi\,,
\end{eqnarray}
where the second line of Eq.~(\ref{vt1}) expresses the reflection symmetry of
the trajectory segments.  In terms of $\phi$ and $l$ we can express
arbitrary averages over the chaotic phase space component as
\begin{eqnarray}\label{trav}
\<\dots\>\ch&=&{\int_{0}^{\pi}\d\phi\,P(\phi)\,\int_{0}^{\Delta l(\phi)}{\d
  l}\,(\dots)\/\int_{0}^{\pi}\d\phi\,P(\phi)\,\int_{0}^{\Delta l(\phi)}\d l}\,. 
\nonumber\\&=&\<\Delta l\>_{\phi}^{-1}
\int_{0}^{\pi}\d\phi\,P(\phi)\,\int_{0}^{\Delta l(\phi)}{\d l}\,(\dots)\,,
\end{eqnarray}
where $\<\dots\>_{\phi}$ is the average over the probability density
$P(\phi)$. Eq.~(\ref{trav}) will be applied in the following sections.

It is useful to note that Eqs.~(\ref{delx}) and (\ref{dell}) allow to
understand the chaotic transport as random walk along the $x$-axis without any
reference to the transversal motion. The discrete steps of this random walk
are $\delx(\phi)$ and the corresponding time increments are $\dell(\phi)/\ve$.
Step $n$ is chosen according to the probability density given in Eq.~(\ref{pcn}).

\subsection{Average chaotic transport velocity}

We will now give an alternative derivation for the chaotic transport velocity
in the case $\sr=0$. We can apply Eq.~(\ref{trav}) to Eq.~(\ref{ergo}) and
simplify the result using the identity
\begin{equation}\label{icphi}
\int_{0}^{\Delta l(\phi)}\d l\,\cos\varphi(\phi,l)=\Delta x(\phi)
\end{equation}
which follows from Eq.~(\ref{vt1}) and should also be obvious from the
geometrical meaning of $\Delta x$ and $\Delta l$. In this way we obtain
\begin{eqnarray}\label{vch2}
v\ch&=&\ve\<\Delta l\>_{\phi}^{-1}\,
\int_{0}^{\pi}\d\phi\,P(\phi)\,\int_{0}^{\Delta l(\phi)}{\d
  l}\,\cos\varphi
\nonumber\\&=&\ve\<\Delta l\>_{\phi}^{-1}\,
\int_{0}^{\pi}\d\phi\,P(\phi)\,\Delta x(\phi)
\nonumber\\
&=&\ve{\<\Delta x\>_{\phi}\over \<\Delta l\>_{\phi}}\,.
\end{eqnarray}
This representation is very intuitive from the random-walk point of view: a
long trajectory consists of $N\to\infty$ steps distributed according to
$P(\phi)$. Therfore the distance between the end points of the trajectory is
$\Delta X=N\<\Delta x\>_{\phi}$ while the total time increment is $\Delta T=N\<\Delta
l\>_{\phi}/\ve$. The average longitudinal component of the velocity is then
$\Delta X/\Delta T$ which is equivalent to Eq.~(\ref{vch2}).
Explicit averaging over $\phi$ yields 
\begin{eqnarray}\label{delxm}
\<\Delta x\>_{\phi}&=&\rc\int_{0}^{\pi}\d\phi\,\sin\phi\,(\sin\phi-\sin\phi')
\nonumber\\&=&\rc\int_{-1}^{+1}\d c\,\sqrt{1-c^2}-r\int_{-1}^{1-b/r}\d c\,\sqrt{1-[c+b/r]^2})
\nonumber\\&=&{\pi\over 2}\rc-{\rc\over 4}\[\pi+2\(1-{b\/\rc}\)\sqrt{{b\/\rc}\(2-{b\/\rc}\)}-2\arcsin\({b\/\rc}-1\)\]
\nonumber\\&=&{r\over 2}(\theta-\sin\theta\cos\theta)\,.
\end{eqnarray}
In agreement with Eq.~(\ref{theta}) we have set $\cos\theta=1-b/r$ in order to
obtain the last line. For the average length of a segment we have
\\
\begin{eqnarray}\label{dellm}
\<\Delta l\>_{\phi}&=&\rc\int_{0}^{\pi}\d\phi\,\sin\phi\,(\phi-\phi') 
\nonumber\\&=&\rc\int_{-1}^{+1}\d c\,\arccos(c)-\rc\int_{-1}^{1-b/r}\d c\,\arccos(c+b/r)
\nonumber\\&=&\pi\rc-{\rc\over 2}\[\(1-{b\/r}\)\(\pi-2\arcsin\({b\/r}-1\)\)+2\sqrt{{b\/r}\(2-{b\/r}\)}\]
\nonumber\\&=&\rc\[\pi{b\/r}-\(\sin\theta-\theta\cos\theta\)\]\,.
\end{eqnarray}
Combining Eqs.~(\ref{vch2})-(\ref{dellm}) we find the expected
result, namely Eq.~(\ref{vch}) with $\sr=0$ (and arbitrary $a$)
\begin{eqnarray}\label{vchr}
v\ch&=&{\ve\/2}{\theta-\sin\theta\cos\theta\/(\pi b/\rc)-
(\sin\theta-\theta\,\cos\theta)}
\\ 
&\approx&{2\ve\/3\pi}\,\sqrt{2b\/\rc}\qquad(\rc\to\infty)\,.
\end{eqnarray}

\begin{figure}[tb]
 \centerline{
  \psfig{figure=\fig/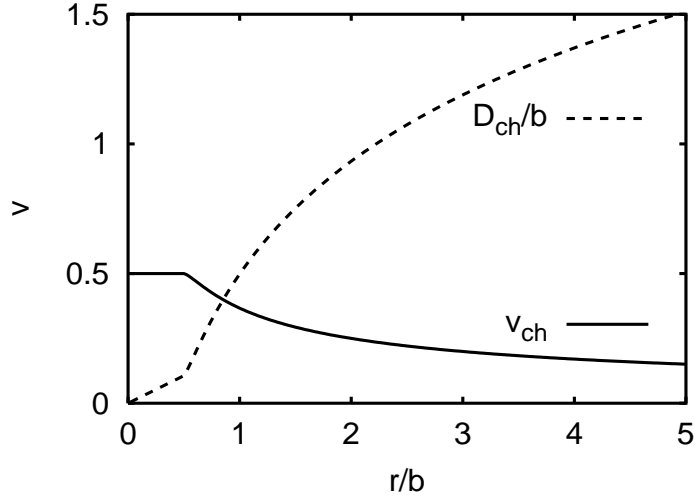,width=10cm}
 }
 \caption{\label{vdch} Dependence of the chaotic transport velocity and
  the diffusion constant on the cyclotron radius $\rc$ for random scattering
  from the lower wall ($\alpha=0$).}
\end{figure}
Fig.~\ref{vdch} shows Eq.~(\ref{vchr}) as a function of the cyclotron radius.
We observe that for small $\rc$ the properties of the system change
qualitatively.  In particular the chaotic transport velocity is constant,
$v\ch=\ve/2$ for $\rc\le b/2$. In this regime a third phase space component of
pinned cyclotron orbits appears in addition to regular skipping and chaotic
trajectories. Hence, the phase-space sum rule for $v\ch$ must be modified and
Eq.~(\ref{vch}) is not valid. In contrast, Eq.~(\ref{vch2}) still applies, but now we
have $\phi'=0$ for all $\phi$ since no chaotic trajectory reaches the upper wall.
Therefore we have
\begin{eqnarray}
\<\Delta x\>_{\phi}&=&2\rc\<\sin\phi\>_{\phi} 
\nonumber\\&=&2\rc{1\over 2}\int_{-1}^{+1}\d\cos\phi\,\sin\phi
\nonumber\\&=&{\pi\/2}\,\rc\,,
\end{eqnarray}
\begin{eqnarray}
\<\Delta l\>_{\phi}&=&2\rc\<\phi\>_{\phi}
\nonumber\\&=&\pi\,\rc
\end{eqnarray}
and finally 
\begin{equation}
v\ch={\ve\/2}\,.
\end{equation}

\subsection{Diffusion constant}

Using the method of the previous subsection we can also calculate
the diffusion constant $D$ of our simplified model. 
Starting point is the well-known Green-Kubo relation expressing $D$ as
integral over the velocity autocorrelation function,
\begin{eqnarray}
D&=&\int_{-\infty}^{+\infty}\d t\,\<\tilde v_{x}(t)\,\tilde v_{x}(0)\>\,.
\end{eqnarray}
Here $\tilde v_{x}(t)=\ve\,\cos\varphi(t)-v\ch $ denotes the fluctuation of
the transport velocity around its mean value and the average $\<\dots\>$ is
taken over all possible trajectories. For the moment we will assume
$\alpha=0$, i.e., no correlations between different segments of a trajectory.  In
this case it suffices to restrict the integral over $t$ to the segment containing
the phase-space point $(\phi_{0},l_{0})$ at $t=0$, that is
to the interval $-{l_{0}}\le\ve t\le \Delta l(\phi_{0})-l_{0}$. We use Eq.~(\ref{trav}) to
average over $(\phi_{0},l_{0})$, make the substitution $l=l_{0}+\ve t$ and
apply Eq.~(\ref{icphi}). In this way we find
\begin{eqnarray}
D_{\alpha=0}&=&\label{intrepdch}
{1\/\<\Delta l\>_{\phi}}
\int_{0}^{\pi}\d\phi_{0}\,P(\phi_{0})
\int_{0}^{\Delta l(\phi_{0})}\d l_{0}\,\d l\,\ve^{-1}
\\&&\qquad\times
\(\ve\cos\varphi(\phi_{0},l)-v\ch\)\(\ve\cos(\phi_{0},l_{0})-v\ch\)
\nonumber\\&=&
{\ve\/\<\Delta l\>_{\phi}}
\int_{0}^{\pi}\d\phi_{0}\,P(\phi_{0})\,
\(\Delta x(\phi_{0})-{v\ch\/\ve}\,\Delta l(\phi_{0})\)^2
\nonumber\\&=&
\ve{\<\[{\Delta x}-(v\ch/\ve)\,\Delta l\]^2\>_{\phi}\over 
\<\Delta l\>_{\phi}}\,.
\label{diffconst}
\end{eqnarray}
This result reduces the calculation of the diffusion constant
to a simple 1D quadrature which is numerically straightforward. An explicit
result is available for strong magnetic field 
\begin{eqnarray}
D_{\alpha=0}
&=&{2\ve\,r\/3\pi} \qquad\(\rc\le{1\/2}\)\,,
\end{eqnarray}
which reproduces the initially linear dependence of the diffusion constant on
the cyclotron radius (Fig.~\ref{vdch}). 

\begin{figure}[tb]
\centerline{
\footnotesize\sf
\begin{tabular}{ll}(a)&(b)\\[-8mm]
  \psfig{figure=\fig/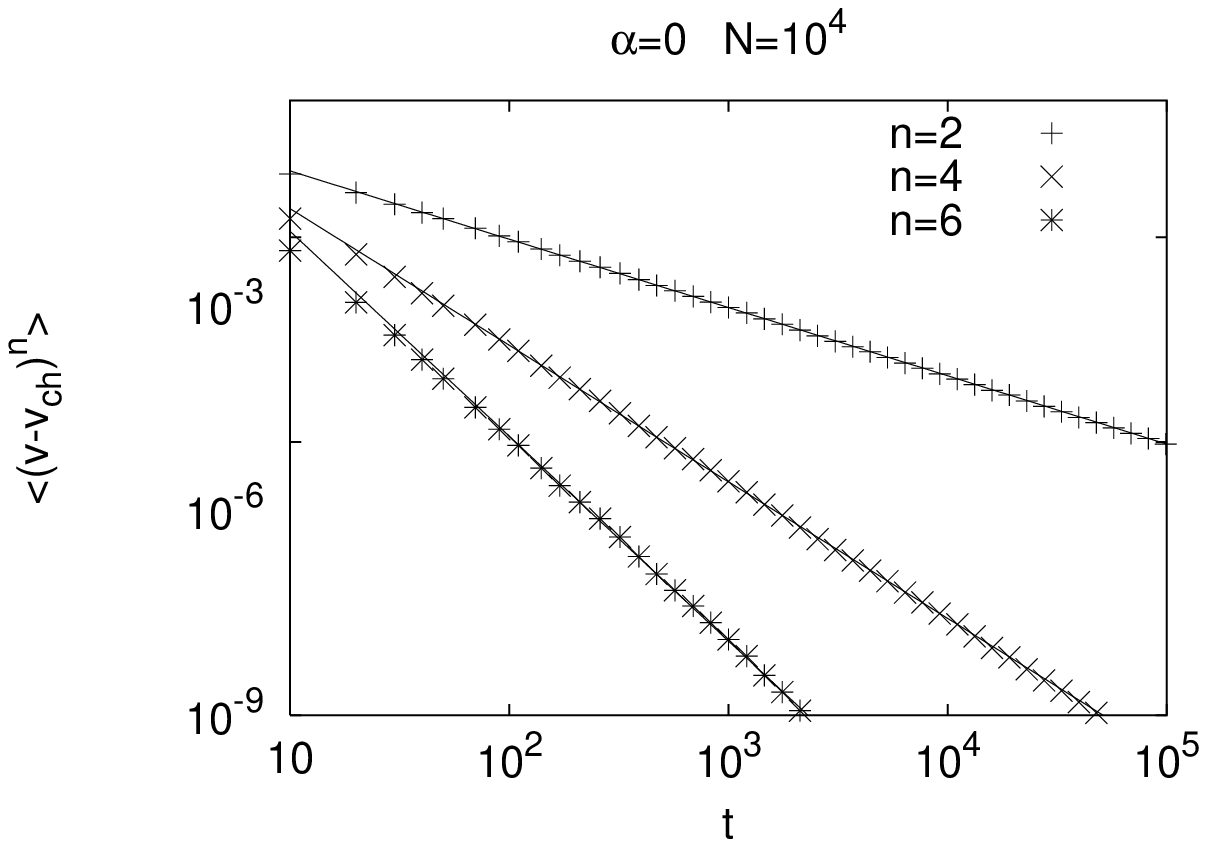,width=6.5cm}&
  \psfig{figure=\fig/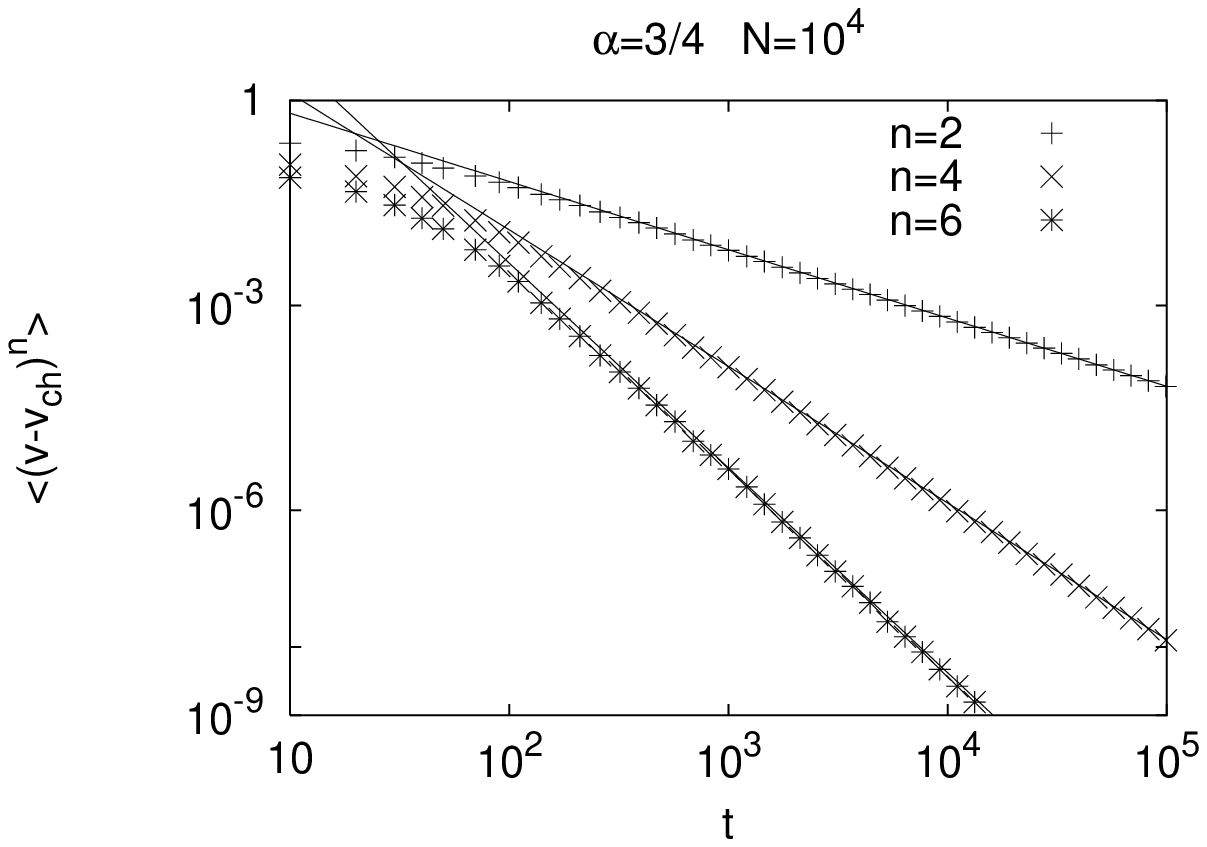,width=6.5cm}
\end{tabular}
}
 \caption{\label{diffuse3} Some higher moments of the distribution of
   $N=10^4$ time-averaged chaotic transport velocities in the magnetic billiard with
   rough lower wall and $\rc=2b$. The probability of specular reflection is
   (a) $\alpha=0$ and (b) $\alpha=3/4$. Straight lines indicate diffusive spreading
   with the diffusion constant given by Eqs.~(\ref{diffconst}) and
   (\ref{dalpha}). The latter correctly predicts that $D\ch$ is greater
   by a factor $7$ in (b).}
\end{figure}

If $\alpha\ne 0$, several consecutive segments of a trajectory can be identical.
Accordingly, the section of the trajectory which is correlated with the point
$(\phi_{0},l_{0})$ is longer and the integration over $l$ in
Eq.~(\ref{intrepdch}) extends over the range $-{l_{0}}-\mu\Delta
l(\phi_{0})\le l\le \Delta l(\phi_{0})-l_{0}+\nu\Delta l(\phi_{0})$ where
$\mu,\nu\ge 0$ count the identical segments before and after the current one.
Hence the integral increases by a factor $\mu+\nu+1$. For $\nu$ identical segments
following the current one we need $\nu$ specular reflections from the lower
wall and finally one random reflection. The probability for this is
$\alpha^{\nu}(1-\alpha)$. Similarly, the probability for $\mu$ identical segments
preceeding the current one is $\alpha^{\mu}(1-\alpha)$. Averaging with these
probabilities over $\mu$, $\nu$ yields
\begin{eqnarray}\label{dalpha}
D_{\alpha}&=&\sum_{\mu,\nu=0}^{\infty}\alpha^{\mu+\nu}\,(1-\alpha)^2\,(\mu+\nu+1)\,D_{0}
\nonumber\\&=&
{1+\alpha\over 1-\alpha}\,D_{0}\,. 
\end{eqnarray}
In Fig.~\ref{diffuse3} we verify these results on the chaotic diffusion constant
numerically.

\section{Summary and discussion}\label{sec:sum}

In this paper we have introduced and studied a particular billiard model which
shows directed chaos. The most appealing feature of this model is the fact
that the phase space has, to a good approximation, a very simple structure
which can be understood both intuitively and analytically. It consists of two
invariant manifolds, one chaotic and one regular. The latter is formed by
trajectories skipping periodically along one perfectly straight wall of a
channel. The other wall has semicircular obstacles or is disordered or rough.
This leads to strong back scattering and thus to a chaotic or random
phase-space component. Both, the particular phase-space structure and the fact that
time-reversal symmetry is broken, are dependent on the presence of a
transversal magnetic field acting on charged particles such as electrons. The
magnetic field should have a suitable intermediate strength.  For very small
magnetic field the phase-space volume of the regular skipping trajectories
goes to zero and, accordingly, the chaotic transport vanishes.  For strong
magnetic field the phase-space structure changes as more and more regular
regions appear.  Finally, for very strong magnetic field, pinned cyclotron
orbits cover most of the phase space.  In this regime the chaotic trajectories
are skipping along the lower wall and $v\ch$ approaches a non-zero constant,
but at the same time the chaotic fraction of phase space goes to zero.

We have stressed in the introduction the analogies between directed chaos in
driven systems and in magnetic billiards.  However, there exists also an
important difference. It has been shown that driven Hamiltonian systems with
directed chaos can generate directed transport in ensembles of particles which
have initially a thermal distribution in phase space
\cite{FYZ00,S+01,D+02,SDK05,M+02,J+05}. They do so without an external
bias force and under the influence of a potential which is periodic both in
space and in time. This is very similar to the concept of stochastic ratchets
(Brownian motors) \cite{Feynman,Rei02,AH02,JAP97}.  Hence the term Hamiltonian
ratchet is appropriate for the models investigated in
\cite{FYZ00,S+01,D+02,SDK05,M+02} and also for the atom-optics setup
studied experimentally in \cite{J+05}.  However, magnetic billiards with
directed chaos are no Hamiltonian ratchets and cannot generate directed
transport from thermal ensembles. The reason is the conservation of energy. We
have seen that the total transport from all regular and chaotic phase-space
components of an energy shell must vanish.  Therefore directed transport in
magnetic billiards requires control over the initial conditions beyond
prescribing a certain distribution of energy values.  For example, electrons
could be placed selectively into the chaotic phase-space component of our
billiard chain if they enter the system from a lead which is attached to the
lower wall of the waveguide.

An interesting extension of the present work would be the inclusion of quantum
effects. Definitely they play an important role when applications to
electronic transport in semiconductor nanostructures are the goal. For a given
geometry and a given time there is always a semiclassical regime in which the
effective value of Planck's constant is sufficiently small such that the
classical results obtained above remain valid. On the other hand, for a given finite
$\hbar$ there should be interesting deviations from our predictions for large
times when tunneling between the regular and the chaotic regions of phase
space may become important and when the presence or absence of disorder in the
system is crucial.

\ack
We profitted from discussions with T. Dittrich and from financial support of
the Volkswagen Foundation (contract I/78235).
\section*{References}

\end{document}